\documentclass[graybox]{svmult}

\usepackage[latin1]{inputenc}
\usepackage[bottom]{footmisc}
\usepackage[all]{xy}
\usepackage{breakurl}
\usepackage{amsmath}
\usepackage{amsfonts}
\usepackage{amssymb}
\usepackage{mathptmx}       
\usepackage{helvet}         
\usepackage{courier}        
\usepackage{type1cm}        
\usepackage{natbib}
\usepackage{graphicx}        
\usepackage{multicol}       
\usepackage[firstpage]{draftwatermark}

\newtheorem{te}{Theorem}[section]
\newtheorem{la}{Lemma}[section]
\newtheorem{co}{Corolary}[section]
\newtheorem{po}{Proposition}[section]
\newtheorem{de}{Definition}[section]
\newtheorem{ex}{Example}[section]
\newtheorem{re}{Remark}[section]
\newtheorem{no}{Notation}[section]

\newcommand{\bpr}{\begin{proof}}
\newcommand{\epr}{\end{proof}}
\newcommand{\bte}{\begin{te}}
\newcommand{\ete}{\end{te}}
\newcommand{\bla}{\begin{la}}
\newcommand{\ela}{\end{la}}
\newcommand{\bco}{\begin{co}}
\newcommand{\eco}{\end{co}}
\newcommand{\bpo}{\begin{po}}
\newcommand{\epo}{\end{po}}
\newcommand{\bde}{\begin{de}}
\newcommand{\ede}{\end{de}}
\newcommand{\bex}{\begin{ex}}
\newcommand{\eex}{\end{ex}}
\newcommand{\bre}{\begin{re}}
\newcommand{\ere}{\end{re}}
\newcommand{\bno}{\begin{no}}
\newcommand{\eno}{\end{no}}

\begin{document}

\title*{Semantic information and artificial intelligence}

\titlerunning{Semantic information and artificial intelligence}

\author{Anderson de Ara\'ujo}

\authorrunning{Ara\'ujo, Anderson de}

\institute{Anderson de Ara\'ujo \at Federal University of ABC (UFABC), Center for Natural and Human Sciences (CCNH), S\~ao Bernardo do Campo, SP, Brazil, \email{anderson.araujo@ufabc.edu.br}}

\maketitle

\vspace*{-2cm}

\abstract{For a computational system to be intelligent, it should be able to perform, at least, basic deductions. Nonetheless, since deductions are, in some sense, equivalent to tautologies, it seems that they do not provide new information. The present article proposes a measure the degree of semantic informativity of valid deductions in a dynamic setting. Concepts of coherency and relevancy, displayed in terms of insertions and deletions on databases, are used to define
semantic informativity. In this way, the article shows that a solution to the problem about the informativity of deductions provides a heuristic principle to improve the deductive power of computational systems.}

\section{Introduction}

For Aristotle, ``every belief comes either through syllogism or from
induction'' (\cite{aristotle1989}). From that, we can infer that every computational
system that aspires to exhibit characteristics of intelligence needs
to have deductive as well as inductive abilities. With respect to
the latter, there are theories that explain why induction is important
for artificial intelligence; for instance, Valiant's probably approximately
correct semantics of learning (\cite{valiant1984,valiant2008}). Nonetheless, as far as
the former is concerned, we have a problem first observed by Hintikka (\cite{hintikka1973}), which can be stated in the following way:
\begin{enumerate}
\item A deduction is valid if, and only if, the conjunction of its premisses,
says $\phi_{1},\ldots,\phi_{n}$, implies its conclusion, $\psi$. 
\item In this case, $\phi_{1}\wedge\cdots\wedge\phi_{n}\to\psi$ is a tautology,
i.e., valid deductions are equivalent to propositions without information. 
\item Therefore, deductions are uninformative.
\end{enumerate}

This was called by Hintikka the \emph{scandal of deduction}. This is a scandal not only because it contradicts Aristotle's maxim that deductions are important for obtaining beliefs, but, mainly, in virtue of the fact that we actually obtain information via deductions. For this and other reasons, Floridi has proposed a theory of strong semantic information in which
semantic information is true well-formed data (Cf. \cite{floridi2004}). From this standpoint, Floridi is capable of explaining why some logical formulas are more informative than others. If we want to explain why deductions, not only propositions, are important for knowledge acquisition and intelligent processing, we cannot, however, apply Floridi's theory. The main reason is that it was designed to measure the static semantic information of the data expressed by
propositions. In contrast, knowledge acquisition and intelligence are dynamic phenomena, associated in some way to the flow of information.

In the present work, we propose to overcome that limitation by defining a measure of semantic information in Floridi's sense, but in the context of a dynamic perspective about the logical features of databases associated
to valid deductions (Section 2). We will restrict ourselves to first-order deductions and will adopt a semantic perspective about them, which means that deductions will be analyzed in terms of structures. There is two reasons for that
choice. The first is that the scandal of deduction is usually conceived in terms of structures associated to valid deductions (Cf. \cite{sequoiah2008}). In other words, we have a problem with the semantic informativity of deductions. The second reason is technical: databases are finite structures and so there is, in general, no complete deductive first-order logical system for finite structures (Cf. \cite{ebbinghaus1999}). 

Besides that, we impose to ourselves the methodological constraint that a good approach of semantic informativity should be robust enough to be applicable to real computational systems. More specifically, we will look for a solution that enables to link semantic information and artificial intelligence. Because of that, we propose to measure the degree of semantic informativity of deductions as a dynamic phenomenon, based on certain explicit definitions of insertions and deletions
on databases. In that context, the concepts of coherency and relevancy will be explained
by the operations of insertion and deletion (Section 3), and so semantic informativity will be conceived in terms of relevancy and coherency (Section 4). This approach leads us to a solution the scandal of deduction (Section 4). Moreover, using straightforward definitions, we shows that our definition of semantic informativity provides an
immediate heuristic principle to improve the deductive power of computational systems in semantic terms.

\section{Databases}

We intend to analyze the semantic informativity obtained via deductions. According to Floridi, semantic information is true well-defined data (\cite{floridi2011}). As far as logic is concerned, we can say that data is in some way expressed by propositions. In general, deductions are compounded of two or more propositions. Thus, we need indeed to consider databases,
because they are just organized collections of data (Cf. \cite{kroenke2007}). From a logical point of view, the usual notion of database can, in turn, be understood in terms of the mathematical concept of structure.

\bde

A \emph{database} is a pair $D=(A,T)$ where $A$ is a finite first-order structure over a signature $S$ and $T$ is a correct (all propositions in $T$ are true in $A$) finite first-order theory about $A$.

\ede

\bex\label{ex1}

Let $D_{1}=(A_{1},T)$ be a database with signature $S=(\{s,l,a\},\{C,E,H\})$,
for $s=\ulcorner\mbox{São\ Paulo}\urcorner$, $l=\ulcorner\mbox{London}\urcorner$,
$a=\ulcorner\mbox{Avenida\ Paulista}\urcorner$, $C=\ulcorner\mbox{City}\urcorner$,
$E=\ulcorner\mbox{Street}\urcorner$ and $H=\ulcorner\mbox{To\ have}\urcorner$,
such that $A_{1}=(\{\bar{s},\bar{l},\bar{a}\},\bar{s}_{s}\bar{l}_{l},\bar{a}_{a},\{\bar{s},\bar{l}\}_{C},\{\bar{a}\}_{E},\{(\bar{s},\bar{a}),(\bar{l},\bar{a})\}_{H})$
and $T=\{\forall x(Cx\to\exists yHxy),\forall x(Cx\vee Ex),\neg El,Cs\}$.

\eex

\bre

In the example \ref{ex1} we have used $\ulcorner\alpha\urcorner=\beta$
to mean that the symbol $\beta$ is a formal representation of the
expression $\alpha$. Besides, $X_{\beta}$ is the interpretation
of $\beta$ in the structure $A$ and we use a bar above letters to
indicate individuals of the domain of $A$.

\ere

The fact that theory $T$ is correct with respect to $A$ does not
exclude, however, the possibility that our database does not correspond
to reality. In the example \ref{ex1}, it is true in $A$ that $Hla\wedge Ea$,
in words, it is true in $A$ that London has a city called Avenida
Paulista, which, until date of the present paper, it is not true.
The theory $T$ represents the fundamental facts of the database that
are took as true, that is to say, they are the \emph{beliefs} of the
database. It is important to observe that $T$ may not be complete
about $A$, it is possible that not all true propositions about $A$
are in $T$; example \ref{ex1} shows this.

We turn now to the dynamics of changes in databases that will permit
us to measure semantic informativity. We propose that these changes
in the structure of databases should preserve the truth proposition
of their theories through operations that we call \emph{structural
operations}. The first structural operation consists in put possibly
new objects in the structure of the given database and to interpret
a possibly new symbols in terms of these objects.

\bde

Let $D=(A,T)$ be a database over a signature $S$. An \emph{insertion}
of the $n$-ary symbol $\sigma\in S'$ in $D$ is a database $D'=(A',T)$
where $A'$ is an structure over $S'=S\cup\{\sigma\}$ with the following
properties: 
\begin{enumerate}
\item $A'(\tau)=A(\tau)$ for all $\tau\neq\sigma$ such that $\tau\in S$; 
\item If $n=0$, then $A'=A\cup\{a\}$ and $A'(\sigma)=a$, provided that,
for all $\phi\in T$, $A'\vDash\phi$; 
\item If $n>0$, then $A'=A\cup\{a_{1},\ldots,a_{n}\}$ and $A'(\sigma)=A(\sigma)\cup\{(a_{1},\ldots,a_{n})\}$,
provided that, for all $\phi\in T$, $A'\vDash\phi$. 
\end{enumerate}
\ede

\bex\label{ex2}

Let $D_{1}=(A,T)$ be the database of the example \ref{ex1}. The
database $D_{2}=(A_{2},T)$ with signature $S'=S\cup\{b\}$, where
$b=\ulcorner\mbox{Shaftesbury\ Avenue}\urcorner$, and $A_{2}=(\{\bar{s},\bar{l},\bar{a}\},\bar{s}_{s},\bar{l}_{l},\bar{a}_{a},\bar{a}_{b},\{\bar{s},\bar{l}\}_{C},\{\bar{a}\}_{E},\{(\bar{s},\bar{a}),(\bar{l},\bar{a})\}_{H})$
is an insertion of $b$ in $D$. On the other hand, $D_{3}=(A_{3},T)$
is an insertion of $E$ in $D_{3}$ where $A_{3}=(\{\bar{s},\bar{l},\bar{a},\bar{b}\},\bar{s}_{s},\bar{l}_{l},\bar{a}_{a},\bar{a}_{b},\{\bar{s},\bar{l}\}_{C},\{\bar{a},\bar{b}\}_{E},\{(\bar{s},\bar{a}),(\bar{l},\bar{a})\}_{H})$
is an $S'$-structure. Nonetheless, for $A^{*}=(\{\bar{s},\bar{l},\bar{a},\bar{b}\},\bar{s}_{s},\bar{l}_{l},\bar{a}_{a},\bar{b}_{b},\{\bar{s},\bar{l}\}_{C},\{\bar{a}\}_{E},\{(\bar{s},\bar{a}),(\bar{l},\bar{a})\}_{H})$,
an $S'$-structure, we have that $D^{*}=(A^{*},T)$ is not an insertion
of $b$ in $D_{1}$ because in this case $A^{*}\nvDash\forall x(Cx\vee Ex)$.

\eex

The example \ref{ex2} shows that it is not necessary to introduce
a new object in the structure of the database to make an insertion
(Cf. database $D_{2}$), it is sufficient to add a possibly new element
in the interpretation of some symbol. On the other hand, it also shows
that it is not sufficient to introduce a new object in the structure
of the database to make an insertion (Cf. database $D^{*}$), it is
necessary to guarantee that the beliefs of the database are still
true in the new structure.

The second structural operation consists in removing possibly old
objects in the structure of the database and to interpret a possibly
new symbol in terms of the remaining objects in the database.

\bde

Let $D=(A,T)$ be a database over a signature $S$. A \emph{deletion}
of the $n$-ary symbol $\sigma\in S'$, $S-\{\sigma\}\subseteq S'\subseteq S$,
from $D$ is a database $D'=(A',T)$ where $A'$ is an structure over
$S'$ with the following properties: 
\begin{enumerate}
\item $A'(\tau)=A(\tau)$ for all $\tau\neq\sigma$ such that $\tau\in S$; 
\item If $n=0$, $A-\{A(\sigma)\}\subseteq A'\subseteq A$ and $A'(\sigma)\in A'$,
provided that, for all $\phi\in T$, $A'\vDash\phi$; 
\item If $n>0$, $A-\{a_{1},\ldots,a_{n}\}\subseteq A'\subseteq A$ and
$A'(\sigma)=A(\sigma)-\{(a_{1},\ldots,a_{n})\}$, provided that, for
all $\phi\in T$, $A'\vDash\phi$. 
\end{enumerate}
\ede

\bex\label{ex3}

Let $D_{1}=(A,T)$ be the database of the example \ref{ex1}. The
database $D'_{2}=(A'_{2},T)$ with signature $S$ and $A'_{2}=(\{\bar{s},\bar{l},\bar{a}\},\bar{a}_{s},\bar{l}_{l},\bar{a}_{a},\{\bar{s},\bar{l}\}_{C},\{\bar{a}\}_{E},\{(\bar{s},\bar{a}),(\bar{l},\bar{a})\}_{H})$
is a deletion of $s$ in $D$. On the other hand, $D'_{3}=(A'_{3},T)$
is a deletion of $H$ in $D'_{2}$ where $A'_{3}=(\{\bar{s},\bar{l},\bar{a}\},\bar{a}_{s},\bar{l}_{l},\bar{a}_{a},\{\bar{s},\bar{l}\}_{C},\{\bar{a}\}_{E},\{(\bar{l},\bar{a})\}_{H})$
is a $S$-structure. Nonetheless, for $A'_{4}=(\{\bar{l},\bar{a}\},\bar{a}_{s},\bar{l}_{l},\bar{a}_{a},\{\bar{l}\}_{C},\{\bar{a}\}_{E},\{(\bar{l},\bar{a})\}_{H})$,
an structure over the signature $S$, we have that $D'_{4}=(A_{4},T)$
is not a deletion of $C$ in $D'_{2}$ because in this case, in despite
of $A'_{4}\vDash\phi$ for $\phi\in T$, we have that $D'_{4}(H)\neq D'_{2}(H)$.
Note, however, that $D'_{4}$ is a deletion of $C$ in $D'_{3}$.

\eex

Example \ref{ex3} illustrates that the restriction $A-\{a_{1},\ldots,a_{n}\}\subseteq A'\subseteq A$
means that we can delete at most just the elements of the domain that
we remove in some way from the interpretation of the symbol that we
are considering in the deletion.

Insertions and deletions on databases are well known primitive operations
(Cf. \cite{kroenke2007}). Nevertheless, to the best of our knowledge, they have being
conceived as undefined notion. Here we have proposed, however, a logical
conception about databases and we have defined explicitly the operations
of insertion and deletion sufficient in order to analyze the importance of semantic information to artificial intelligence. In \cite{araujo2014b}, a more strict notion of structural operation is given.

\section{Coherency and relevancy}

In this section, we propose a dynamic perspective about coherency
and relevancy. This approach will permit us to evaluate how many structural
operations a proposition requires to become true. We will use these
concepts to define the semantic informativity in the next section.

\bde

An \emph{update $\bar{D}$} of an $S$-database $D$ is a finite or
infinite sequence $\bar{D}=(D_{i}:0<i\leq\omega)$ where $D_{1}=D$
and each $D_{i+1}$ is a insertion or deletion in $D_{i}$. An update
$\bar{D}$ of $D$ is \emph{coherent }with a proposition $\phi$ if
$\bar{D}=(D_{1},D_{2},\ldots,D_{n})$ and $A_{n}\vDash\phi$; otherwise, $\bar{D}$ is said to
be \emph{incoherent} with $\phi$.

\ede

\bex\label{ex4}

Let $D_{1}$ be the database of the example \ref{ex1} and $D{}_{2}$
be the databases of the example \ref{ex2}. The sequence $\bar{D}=(D_{1},D{}_{2})$
is an update of $D$ coherent with $Eb$ and $Hlb$. Let $D_{1}$
be the database of the example \ref{ex1} and $D'_{2}$, $D'_{3}$
and $D'_{4}$ be the databases of the example \ref{ex3}. The sequence
$\bar{D}'=(D_{1},D'_{2},D'_{3},D'_{4})$ is an update of $D$ coherent
with $Es\wedge\neg Hsa$ but not with $s=a$ because the last proposition
is false in $A'_{4}=(\{\bar{l},\bar{a}\},\bar{a}_{s},\bar{l}_{l},\bar{a}_{a},\{\bar{l}\}_{C},\{\bar{a}\}_{E},\{(\bar{l},\bar{a})\}_{H})$.

\eex

In other words, an update for a proposition $\phi$ is a sequence
of changes in a given database that produces a structure in which
$\phi$ is true. In this way, we can measure the amount of coherency
of propositions.

\bde\label{dec}

Let $\bar{D}=(D_{1},D_{2},\ldots,D_{n})$ be an update of the database
$D$. If $\bar{D}$ is coherent with $\phi$, we define the \emph{coherency}
of $\phi$ with $\bar{D}$ by 
\[
H_{\bar{D}}(\phi)=\frac{\min\{m\leq n:A_{m}\vDash\phi\}}{\sum_{i=1}^{m}i}
\]

but if $\bar{D}$ is incoherent with $\phi$ , then 
\[
H_{\bar{D}}(\phi)=0.
\]

A proposition $\phi$ is said to be \emph{coherent} with the database
$D$ if $H_{\bar{D}}(\phi)>0$ for some update $\bar{D}$, otherwise,
$\phi$ is \emph{incoherent} with $D$.

\ede

\bre

In the definition of coherency the denominator $\sum_{i=1}^{m}i$
is used in to order to normalize the definition (the coherency is
a non-negative real number smaller than or equal to 1).

\ere

\bex\label{ex5}

The coherence of $Eb$ and $Hlb$ with the update $\bar{D}$ of
the example \ref{ex4} is the same $2/3$, i.e., $H_{\bar{D}}(Eb)=H_{\bar{D}}(Hlb)\approx0.66$
and so $H_{\bar{D}}(Eb\wedge Hlb)=H_{\bar{D}}(Eb\vee Hlb)\approx0.66$.
On the other hand, with respect to the coherence of $Es$, $\neg Hsa$
and $\neg s=a$ and with the update $\bar{D}'$ of the example \ref{ex4},
we have $H_{\bar{D}'}(Es)\approx0.66$, $H_{\bar{D}'}(\neg Hsa)=0.4$,
$H_{\bar{D}'}(s=a)=0$ and so $H_{\bar{D}'}(Es\wedge\neg Hsa)=0.4$
but $H_{\bar{D}'}(Es\wedge s=a)=0$.

\eex

The example \ref{ex5} exhibits that, given an update, we can have
different propositions with different coherency, but we can have
different propositions with the same coherency as well. The fact that
so $H_{\bar{D}}(Eb)=H_{\bar{D}}(Hlb)=H_{\bar{D}}(Eb\wedge Hlb)\approx0.66$
shows that concept of coherency \emph{is not} a measure of the complexity
of propositions. It seems natural to think that $Eb\wedge Hlb$ is
in some sense more complex than $Eb$ and $Hlb$. Here we do not have
this phenomena. Moreover, the fact that $H_{\bar{D}}(Eb\wedge Hlb)=H_{\bar{D}}(Eb\vee Hlb)\approx0.66$
makes clear that, since some propositions have a given coherency,
many others will have the same coherency. Another
interesting point is that $H_{\bar{D}'}(Es)>H_{\bar{D}'}(\neg Hsa)$
but $H_{\bar{D}'}(\neg Hsa)=H_{\bar{D}}(Eb\wedge Hlb)\approx0.33$.
This reflects the fact that updates are sequences. First, we had made $Es$ coherent
with $\bar{D}'$, and later $\neg Hsa$ we made coherent with $\bar{D}$.
When $\neg Hsa$ is coherent with $\bar{D}'$ there is nothing more
to be done, as far as the conjunction $Es\wedge\neg Hsa$ is concerned.

These remarks show that our approach is very different from the one given in (Cf. \cite{dagostino2009}). It is not an analysis of some concept of complexity associated to semantic information. In \cite{araujo2014b}, we do an analysis of informational complexity similar to one presented here about coherency, but this two concepts are different. In further works, we will examine the relation between them. For now, we are interested in artificial intelligence. With respect to that, we can obtain an important result in the direction of a solution to the scandal of deduction.

\bpo

For every database $D=(A,T)$ and update $\bar{D}$ coherent with
$\phi$, $H_{\bar{D}}(\phi)=1$ for every $\phi$ such that $A\vDash\phi$.
In particular, for $\phi$ a tautology in the language of $D$, $H_{\bar{D}}(\phi)=1$,
but if $\phi$ is not in the language of $D$, $0<H_{D}(\phi)<1$.
In contrast, for every contradiction $\psi$ in any language, $H_{\bar{D}}(\psi)=0$
for every update $\bar{D}$ of $D$.

\epo

In virtue of our focus in this paper is conceptual, we will not provide
proofs here (Cf. \cite{araujo2014b}). By now, we only observe that if a tautology has symbols
different from the ones in the language of the database, it will be
necessary to make some changes in order to make that tautology become
true. In contrast, a proposition is incoherent with a database when
there is no way to change it in order to become the proposition true
and, for this reason, contradictions are never coherent.

We turn now to the concept of relevancy. For that, let us introduce
a notation. Consider $(\phi_{1},\phi_{2},\ldots,\phi_{n})$ a valid
deduction of formulas over the signature $S$ whose premisses are
the set $\Gamma=\{\phi_{1},\phi_{2},\ldots,\phi_{m}\}$ and its conclusion
is $\phi=\phi_{n}$. We represent this deduction by $\Gamma\{\phi\}$.

\bde

Let $\bar{D}=(D_{1},\ldots,D_{n})$ be an update of the $S$-database
$D=(A,T)$ coherent with $\phi$. The \emph{relevant premises} of
the deduction $\Gamma\{\phi\}$ with respect to $\bar{D}$ are the
premises that are true in $D_{n}$ but are not logical consequences
of $T$, i.e., the propositions in the set $\bar{D}(\Gamma)$ of all
$\psi\in\Gamma$ for which $D_{n}\vDash\psi$ but $T\nvDash\psi$.

\ede

\bex\label{ex6}

Let $\bar{D}''=(D_{1})$. Then, $\bar{D}''(\{Ea\}\{\exists xEx\})=\{Ea\}$. Now let us consider a more complex example. Let $\bar{D}=(D_{1},D_{2})$ be the update of \ref{ex4}. In this case, $\bar{D}(\{\forall x(Cx\to\neg Ex),Cb\}\{\neg Eb\})$
is not defined because $\neg Eb$ is false in $A_{2}=(\{\bar{s},\bar{l},\bar{a}\},\bar{s}_{s},\bar{l}_{l},\bar{a}_{a},\bar{a}_{b},\{\bar{s},\bar{l}\}_{C},\{\bar{a}\}_{E},\{(\bar{s},\bar{a}),(\bar{l},\bar{a})\}_{H})$. Nonetheless, consider the new update $\bar{D}'''=(D_{1},D_{2},D_{3},D_{4}, D_{5})$ such that $D_{3}$ is the insertion in example \ref{ex2}, $A_{4}=(\{\bar{s},\bar{l},\bar{a},\bar{b}\},\bar{s}_{s},\bar{l}_{l},\bar{a}_{a},\bar{b}_{b},\{\bar{s},\bar{l}\}_{C},\{\bar{a},\bar{b}\}_{E},\{(\bar{s},\bar{a}),(\bar{l},\bar{a})\}_{H})$ and $A_{5}=(\{\bar{s},\bar{l},\bar{a},\bar{b}\},\bar{s}_{s},\bar{l}_{l},\bar{a}_{a},\bar{b}_{b},\{\bar{s},\bar{l}\}_{C},\{\bar{a}\}_{E},\{(\bar{s},\bar{a}),(\bar{l},\bar{a})\}_{H})$. Then, $\bar{D}'''(\{\forall x(Cx\to\neg Ex),Cb\}\{\neg Eb\})=\{\forall x(Cx\to\neg Ex)\}$.

\eex

In our definition of relevant premises, we have adopted a semantic
perspective oriented to conclusion of deductions: the relevancy of
the premises of a deductions are determined according to an update
in which its conclusion is true. Example \ref{ex6} illustrates that
point, because it is only possible to evaluate the relevancy of $\{\forall x(Cx\to\neg Ex),Cb\}\{\neg Eb\}$
in an update like $\bar{D}'$ in which the conclusion $\neg Eb$ is
true. Another point to be noted is that we have chosen a strong requirement
about what kind of premises could be relevant: the relevant premises
are just the non-logical consequences of our believes.

\bde

Let $D$ be an $S$-database. If $\bar{D}$ is an update of $D$ coherent
with $\phi$, the \emph{relevancy} $R_{\bar{D}}(\Gamma)$ of the deduction
$\Gamma\{\phi\}$ in $\bar{D}$ is the cardinality of $\bar{D}(\Gamma)$
divided by the cardinality of $\Gamma$,i.e., 
\[
R_{\bar{D}}(\Gamma)=\frac{|\bar{D}(\Gamma)|}{|\Gamma|},
\]

but, if $\bar{D}$ is incoherent with $\phi$, then $R_{\bar{D}}(\Gamma)=0$.

\ede

\bex

We have showed in example \ref{ex6} that $R_{\bar{D}''}(\{Ea\}\{\exists xEx\})=1$
and $R_{\bar{D}'''}(\{\forall x(Cx\to\neg Ex),Cb\}\{\neg Eb\})=0.5$.

\eex

In the example above, $R_{\bar{D}'''}(\{\forall x(Cx\to\neg Ex),Cb\}\{\neg Eb\})=0.5$
show us that we can have valid deductions with non-null relevancy
in extended languages. Nonetheless, the fact $R_{\bar{D}''}(\{Ea\}\{\exists xEx\})=1$
shows that is not necessary to consider extended languages to find
deductions with non-null relevancy. Thus, we have a result that will
be central to our solution of the scandal of deduction.

\bpo

For every database $D=(A,T)$, update $\bar{D}$ of $D$ and deduction
$\Gamma\{\phi\}$, if $T$ is a complete theory of $A$ or $\Gamma=\oslash$,
then $R_{\bar{D}}(\Gamma)=0$. In particular, tautologies and contradictions
have null relevancy.

\epo

Therefore, deductions can be relevant only when we do not have a complete
theory of the structure of the database. Moreover, as deductions,
isolated logical facts (tautologies and contradictions) have no relevance.
This means that we have at hand a deductive notion of relevancy.

\section{Semantic informativity and artificial intelligence}

Having at hand the dynamic concepts of coherence and relevance, now
it seems reasonable to say that the more coherent the conclusion of
a valid deduction is the more informative it is, but the more relevant
its premises are the more information they provide. We use this intuition
to define the semantic informativity of valid deductions.

\bde

The \emph{semantic informativity} $I_{\bar{D}}(\Gamma\{\phi\})$ of
a valid deduction $\Gamma\{\phi\}$ in the update $\bar{D}$ of the
database $D$ is defined by 
\[
I_{\bar{D}}(\Gamma\{\phi\})=R_{\bar{D}}(\Gamma)H_{\bar{D}}(\phi).
\]

\ede

The idea behind the definition of semantic informativity of a valid
deduction $\Gamma\{\phi\}$ is that $I_{\bar{D}}(\Gamma\{\phi\})$
is directly proportional to the relevance of its premises $\Gamma$
and to the coherency of its conclusion $\phi$. Given $\Gamma\{\phi\}$
and an update $\bar{D}$ of $D$, if we have $R_{\bar{D}}(\Gamma)=0$
or $H_{\bar{D}}(\phi)=0$, then the semantic informativity of $\Gamma\{\phi\}$
is zero, it does not matter how $\Gamma\{\phi\}$ is. Now, if $H_{\bar{D}}(\phi)=0$,
then, by definition, $R_{\bar{D}}(\Gamma)=0$. Thus, if the computational
system, whose database is $D$, intends to evaluate $I_{\bar{D}}(\Gamma\{\phi\})$
for some update $\bar{D}$, it should look for a $\bar{D}$ coherent
with $\phi$, i.e., a $\bar{D}$ for which $H_{\bar{D}}(\phi)>0$.
In other words, our analysis of the semantic informativity is oriented
to the conclusion of valid deductions - as we did with respect to
relevancy.

\bex\label{ex7}

Given the updates $\bar{D}''$ and $\bar{D}'''$ of the example \ref{ex6}.
Then, $I_{\bar{D}''}(\{Ea\}\{\exists xEx\})=1\cdot1=1$ and $I_{\bar{D}'''}(\{\forall x(Cx\to\neg Ex),Cb\}\{\neg Eb\})=0.5\cdot 5/15 \approx 0.17$.

\eex

In the definition of $I_{\bar{D}}(\Gamma\{\phi\})$ the relevancy
of the premisses, $R_{\bar{D}}(\Gamma)$, is a factor of the coherency
of the conclusion, $H_{\bar{D}}(\phi)$. For that reason, if a computational
systems intends to evaluate the semantic informativity of a proposition
$\phi$, it should measure $H_{\bar{D}}(\phi)$ and, then, multiply
it by its relevancy $R_{\bar{D}}(\{\phi\})$. Hence, the semantic
informativity of a proposition $\phi$ can be conceived as a special
case of the informativity of the valid deduction $\{\phi\}\{\phi\}$.

\bde\label{de}

The \emph{semantic informativity} $I_{\bar{D}}(\phi)$ of a proposition
$\phi$ in the update $\bar{D}$ of the database $D$ is defined by
\[
I_{\bar{D}}(\phi)=I_{\bar{D}}(\{\phi\}\{\phi\}).
\]

\ede

\bex\label{ex8}

Considering the update $\bar{D}''$ of example \ref{ex6}, we have
that $I_{\bar{D}'''}(Ea)=I_{\bar{D}'''}(\exists xEx)=1$ but $I_{\bar{D}'''}(Ea\to\exists xEx)=0$.
If we consider the update $\bar{D}'''$ of example \ref{ex6}, we
have that also have that $I_{\bar{D}'''}((\forall x(Cx\to\neg Ex)\wedge Cb)\to\neg Eb)=0$,
but $I_{\bar{D}'''}(\forall x(Cx\to\neg Ex))=1$, $I_{\bar{D}'''}(Cb)=0$
and $I_{\bar{D}'''}(\neg Eb)=0.4$.

\eex

Example \ref{ex8} shows that semantic informativity measures how
many structural operations we do in order to obtain the semantic information
of a proposition. It is for that reason that $I_{\bar{D}'''}(Cb)=0$,
false well-defined data is not semantically informative; it should
be true. In other words, it is a measure of semantic information in
Floridi's sense (Cf. \cite{floridi2011}). From this, we can solve Hintikka' scandal of deduction.

\bpo

For every valid deduction $\psi_{1},\ldots,\psi_{n}\vDash\phi$ in
the language of $D$, $I_{\bar{D}}((\psi_{1}\wedge\cdots\wedge\psi_{n})\to\phi)=0$
for every update $\bar{D}$. Nonetheless, if $\psi_{1},\ldots,\psi_{n}\vDash\phi$
is not in the language of $D$, $I_{\bar{D}}((\psi_{1}\wedge\cdots\wedge\psi_{n})\to\phi)>0$
for $\bar{D}$ coherent with $(\psi_{1}\wedge\cdots\wedge\psi_{n})\to\phi$.

\epo

This proposition is a solution to the scandal of deduction in two different
senses. First, it shows that we can have an informative valid deduction
$\{\psi_{1},\ldots,\psi_{n}\}\{\phi\}$ whose associated conditional
$\psi_{1},\ldots,\psi_{n}\to\phi$ is uninformative, for example, the one given in example \ref{ex7}. Second, it shows that it is not completely true that tautologies are always uninformative. When we need to interpret new symbols, we have some semantic information, notably, the one sufficient to perceive that we have a true proposition - this is a natural consequence of our approach.

From this standpoint, we are going to make a simple, but important, remark to establish a relationship between semantic informativity and artificial intelligence.

Given an update $\bar{D}=(D_{1},\ldots,D_{n})$ of $D=(A,T)$ and
a deduction $\{\phi\}\{\phi\}$, either $R_{\bar{D}}(\{\phi\})=0$
or $R_{\bar{D}}(\{\phi\})=1$. If $R_{\bar{D}}(\{\phi\})=0$, then
either $T\vDash\phi$ or $D_{n}\nvDash\phi$. If $T\vDash\phi$, then
there is an update $\bar{D}'$ of $D$ such that $H_{\bar{D}'}(\phi)=1$,
notably, $\bar{D}'=(D)$. If $D_{n}\nvDash\phi$, then $H_{\bar{D}'}(\phi)=0$. Finally, if $R_{\bar{D}}(\{\phi\})=1$,
then $T\nvDash\phi$ as well as $D_{n}\vDash\phi$ and so $I_{\bar{D}}(\phi)=H_{\bar{D}}(\phi)>0$.
Therefore, we conclude that the relevancy of a proposition does not determine its coherency. On the other
hand, if $H_{\bar{D}}(\phi)=0$, then $R_{\bar{D}}(\{\phi\})=0$,
but if $H_{\bar{D}}(\phi)>0$, this neither necessarily imply that
either $R_{\bar{D}}(\{\phi\})=1$ nor $R_{\bar{D}}(\{\phi\})=0$,
because this depends whether $T\vDash\phi$. Hence, we also conclude
that the coherency of a proposition does not determine its
relevancy too. Combining this two conclusions we obtain a general conclusion:
the semantic informativity of propositions cannot be determined by
its coherency or relevancy alone. This reinforces our definition of
semantic informativity. It is not an arbitrary definition, in fact
semantic informativity is a relationship between both, coherence and
relevancy. What is the moral for artificial intelligence?

In the studies of pragmatics (a linguistics' area
of research), Wilson and Sperber formulated two principles about relevant
information in human linguistic practice:

\begin{quote}
``Relevance may be assessed in terms of cognitive effects and processing
effort: (a) other things being equal, the greater the positive cognitive
effects achieved by processing an input, the greater the relevance
of the input to the individual at that time; (b) other things being
equal, the greater the processing effort expended, the lower the relevance
of the input to the individual at that time.'' \cite{wilson2004}[p.608]
\end{quote}

Our general conclusion that semantic informativity of propositions
cannot be determined by its coherency or relevancy alone shows that
the two Wilson and Sperber's principles (a) and (b) are in fact parts
of one general principle associated to semantic information. Let us
put that in precise terms.

\bde

The \emph{changes} that a proposition $\phi$ \emph{requires} are
the structural operations, insertions and deletions, that generate
an update $\bar{D}$ of a given database $D=(A,T)$ coherent with
$\phi$. A proposition $\phi$ is \emph{new} if $\phi$ is not false
in $A$ and is not a consequence of the theory $T$ of the database
$D=(A,T)$.

\ede

\bpo\label{p1}

The less changes a new proposition requires, the more informative
it is.

\epo

Proposition \ref{p1} is a direct consequence of definition \ref{de}.
As we have showed that our definition \ref{de} is not, in turn, arbitrary,
this means that \ref{p1} is not arbitrary. \ref{p1} is a reformulation
of the Wilson and Sperber's principle (b) above, but it is important
to note the differences between them. Wilson and Sperber's principle
(b) is an empirical matter under discussion among linguistics (Cf. \cite{wilson2004}). The proposition \ref{p1} is a reformulation of our definition of semantic informativity. Using the same strategy, we can also obtain
a formal version of Wilson and Sperber's principle (a).

\bde

If a valid deduction $\Gamma\{\phi\}$ has non-null relevancy in a
given update $\bar{D}$ and its conclusion $\phi$ is new, then the
\emph{results} that it \emph{produces} are its relevant premisses
and its conclusion, i.e., $\bar{D}(\Gamma)\cup\{\phi\}$, but if $\phi$
is not new, then the \emph{results} that it \emph{produces} are just
its relevant premisses $\bar{D}(\Gamma)$.

\ede

\bpo

The more results a valid deduction produces, the more informative
it is.

\epo

We can, then, combine these two propositions in an schematic one.

\bpo[Principle of semantic informativity]

To increase semantic informativity, an intelligent agent, with respect
to its database, should perform little changes and produce big results.

\epo

In recent works (Cf. \cite{valiant2008}), Valiant have argued that one of the most
important challenges in artificial intelligence is that of understanding
how computational systems that acquire and manipulate commonsense
knowledge can be created. With respect to that point, he explains
that some of the lessons from his PAC theory is this:

\begin{quotation}
``We note that an actual system will attempt to learn many concepts
simultaneously. It will succeed for those for which it has enough
data, and that are simple enough when expressed in terms of the previously
reliably learned concepts that they lie in the learnable class.'' \cite{valiant2008}[p.6]
\end{quotation}

We can read Valiant's perspective in terms of the principle of semantic
informativity. The simple propositions are the more coherent propositions,
the ones that requires little changes in the database. To have enough
data is to have propositions sufficient to deduce another propositions
and this means that deductions with more results are preferable. Of
course, Valiant's remark relies on PAC, a theory about learnability,
not on deductivity. It is necessary to develop further works to make
clear the relationship between this two concepts. Since we have designed
a concept of semantic informativity implementable in real systems,
it seems, however, that the possibility of realizing that is open.

\section{Conclusion}

We have proposed to measure the degree of semantic informativity of
deductions by means of dynamic concepts of relevancy and coherency.
In an schematic form, we can express our approach in the following
way:

\[
\mbox{ Semantic informativity }=\mbox{ Relevancy }\times\mbox{ Coherency.}
\]

In accordance with this conception, we showed how the scandal of deduction can be solved. Our solution is that valid deductions are not always equivalent to propositions without information. It is important to note, however, that this problem is not solve in its totality, because here we have analyzed semantic information only from the point of view of relevancy and coherency. Another crucial concept associated to semantic information is the notion of complexity. In \cite{araujo2014b}, we treat this subject, but a complete analysis of the relation among semantic information, relevance, coherency and complexity is necessary.

In this respect, we have derived a principle of semantic informativity that, when applied to computational intelligent systems, means that an intelligent agent should make few changes in its database and obtain big results. This seems an obvious observation, but it is not. The expressions ``few changes'' and ``big results'' here have a technical sense which opens the possibility of relating semantic information and artificial intelligence in a precise way. Indeed, there is a lot
of possible developments to be explored, we would like to indicate three.

The first one is to investigate the connections between semantic informativity and machine learning, specially, with respect to Valiant's semantic theory of learning (PAC). As in Valiant's PAC, we could establish probability distributions on the possible updates and delineate goals for them - the principle of semantic informativity could play an important role in this point. Moreover, we can also introduce computational complexity constrains to agent semantic informativity. Thus, it will
be possible to analyze how many efficient updates (time and space requirements bounded by a function of the proposition size) are necessary for a given proposition to be coherent with the database. In this way, we will be able, for example, to compare the learnability of the concepts which occur in propositions, in the Valiant's sense (Cf. \cite{valiant1984}), with respect to their semantic information.

The second possible line of research is to develop a complete dynamic theory of the semantic informativity by incorporating belief revision in the line of AGM theory (Cf. \cite{alchourron1985}). In the present
paper, the believes of the database have been maintained fixed, but
a more realistic approach should incorporate revision of believes.
For example, if we consider distributed systems, the agents probably
will have some different beliefs. In this case, it will be necessary
to analyze the changes of semantic information, conflicting data and
so on. 

The last point to be explored, but no less important, is to analyze
the relationship between our dynamic perspective about semantic information
and other static approaches, mainly, with respect to Floridi's theory
of strong semantic information (\cite{floridi2004}). It is important
to observe that we have proposed a kind of hegelian conception about
semantic information, according to which semantic informativity is
analyzed in semantic terms, whereas, for example, Floridi's conception
is kantian, in the sense that it analyzes the relationship between
propositions and the world in order to understand the transcendental
conditions of semantic information (Cf. \cite{floridi2011}). Our
approach seems to be a hegelian turn in the philosophy of information
similar to what Brandom did with respect to the philosophy of language
(Cf. \cite{brandom1989}).

\begin{acknowledgement}

I would like to thank Viviane Beraldo for her encouragement, to Luciano
Floridi for his comments on my talk given at PT-AI2013, to Pedro Carrasqueira and William Steinle for their comments and to an anonymous referee for his (her) criticism on a previous version of this paper. This work was supported by S\~ao Paulo Research Foundation (FAPESP) [2011/07781-2].

\end{acknowledgement}

\bibliography{bibliografia}

\end{document}